\documentclass[aip,jap,amsmath,amssymb,reprint]{revtex4-1}

\usepackage{graphicx}
\usepackage{dcolumn}
\usepackage{bm}
\usepackage{hyperref}

\begin{document}

\title[]{Probing the Magnetodynamics of Magnetic Tunnel Junctions with the Aid of SiGe HBTs}

\author{J. Dark}
 \email{JFDark@gatech.edu}
\affiliation{School of Physics, Georgia Institute of Technology, Atlanta, Georgia 30332, USA}
\author{H. Ying}
\affiliation{School of Electrical and Computer Engineering, Georgia Institute of Technology, Atlanta, Georgia 30332, USA}
\author{G. Nunn}
\affiliation{School of Physics, Georgia Institute of Technology, Atlanta, Georgia 30332, USA}
\author{J. D. Cressler}
\affiliation{School of Electrical and Computer Engineering, Georgia Institute of Technology, Atlanta, Georgia 30332, USA}
\author{D. Davidovi\'c}
\affiliation{School of Physics, Georgia Institute of Technology, Atlanta, Georgia 30332, USA}

\date{\today}

\begin{abstract}
High impedance ($\sim$1~M$\Omega$) magnetic tunnel junctions (MTJs) are used to observe and record the magnetodynamics of the nanomagnets that form the junctions themselves. To counteract the bandwidth limitations caused by the high impedance of the junction and the parasitic capacitance intrinsic to any cryogenic system, silicon-germanium heterojunction bipolar transistors (SiGe HBTs) are used as cryogenic preamplifiers for the MTJs. The resulting measurement improvements include an increase in bandwidth by a factor of 3.89, an increase in signal-to-noise ratio by a factor of 6.62, and a gain of 7.75 of the TMR signal produced by the MTJ. The limitation to the measurement system was found to be from the external, room temperature electronics. Despite this limitation, these improvements allow for better time-resolved magnetodynamics measurements of the MTJs. These experiments pave the way for future cryogenic, magnetodynamics measurement improvements, and could even be useful in cryogenic memory applications.
\end{abstract}

\maketitle
\section{\label{intro}Introduction}
As technology decreases in size and power consumption and increases in speed, the magnetic systems used for memory storage in such technologies need to advance in the same directions. In the interest of size, the individual magnetic elements have grown smaller over the years\cite{Doyle1998}. Work in the field of spin transfer torque has led to lower power consumption\cite{Aradhya2016}. For overall speed, the switching mechanism of nanomagnets directly affects how quickly the magnets will undergo magnetic reversal and stabilize in a different orientaion\cite{Xiao2006}. Consequently, the magnetodynamics of a single nanomagnet leads to a speed limitation on any memory device built using the nanomagnet as its basic memory element.

The switching process in a magnet is a transport of the magnetic moment from an initial high energy metastable state to a final equilibrium state. Therefore, near equilibrium techniques such as ferromagnetic resonance cannot easily access the switching dynamics. However, there has been success in studying the magnetodynamics of switching in nanomagnets by averaging effects over large arrays of particles and by measuring single particles locally. Local studies of the dynamics make use of the magneto-optic Kerr effect, superconducting quantum interference devices, various forms of microscopy, and other electron transport techniques \cite{Martin2003}. With electron transport, magnetoresistance is exploited in various ways ultimately boiling down to Slonczewski's theories involving spin-dependent transport\cite{Slonczewski1989}. Some devices used include magnetic tunnel junctions (MTJs)\cite{Koch1998, Hahn2016}, spin valves \cite{Koch2004}, and nanoparticles embedded inside a tunnel junction \cite{Birk2010,Jiang2013,Gartland2015,Gartland20152,Jiang2016,Jiang2017}. Previous studies of switching dynamics in MTJs using electron transport have used low impedance samples at room temperature\cite{Krivorotov2005,Aoki2010}. For this article, high impedance MTJs ($\sim$1~M$\Omega$) are studied at cryogenic temperatures. The higher impedance MTJs and low temperature environment allow for future use of the same measurement system to study ferromagnetic quantum dots at higher bandwidths\cite{Jiang2017}.

It has long been known that the tunnel magnetoresistance (TMR) of MTJs increases with decreasing temperature \cite{Yaoi1993}. As a result, cooling the MTJs in a cryostat down to 8K has the added benefit of a larger signal resulting in better resolution of the dynamics of the system. On the other hand, measuring fast moving signals inside a cryostat brings about other difficulties. Long signal wires give rise to unwanted parasitic capacitance. This parasitic capacitance coupled to the large resistance of the MTJ leads to bandwidth limitations on any measurement.

Successful improvements have been implemented in similar situations for qubit readout\cite{Tracy2016}, charge detection with a quantum point contact\cite{Vink2007}, and single-electron transistor operation\cite{Pettersson1996,Visscher1996,Curry2015}. Although other solutions exist, those implemented in the above references involve directly interfacing an amplifier with the device under test inside the cryostat to effectively transform the impedance of the device and increase the bandwidth of the overall measurement. This approach also has the potential to amplify the initial signal before it exits the cryostat, which can lead to an increased signal-to-noise ratio (SNR).

SiGe HBTs have recently been shown to operate at milliKelvin temperatures\cite{Ying2017}, and a theory of their cryogenic operation has begun to develop\cite{Davidovic2017,Ying2018}. Although other technologies can also operate at cryogenic temperatures\cite{Pospieszalski2005}, SiGe HBTs have a better scaling potential\cite{Cressler2010}. This is not the focus of the article, but forward thinking  encourages the use of a technology that would be compatible for future applications, such as cryogenic memory. Furthermore, as proven by Curry et al., "off-the-shelf" SiGe HBTs can be used for cryogenics measurements\cite{Curry2015}. As a result, a SiGe HBT is used to combat the bandwidth limitations imposed by the cryostat signal wires and large MTJ resistance.

\begin{figure}
\setlength{\abovecaptionskip}{5pt}
\setlength{\belowcaptionskip}{-10pt}
\includegraphics[scale=.52]{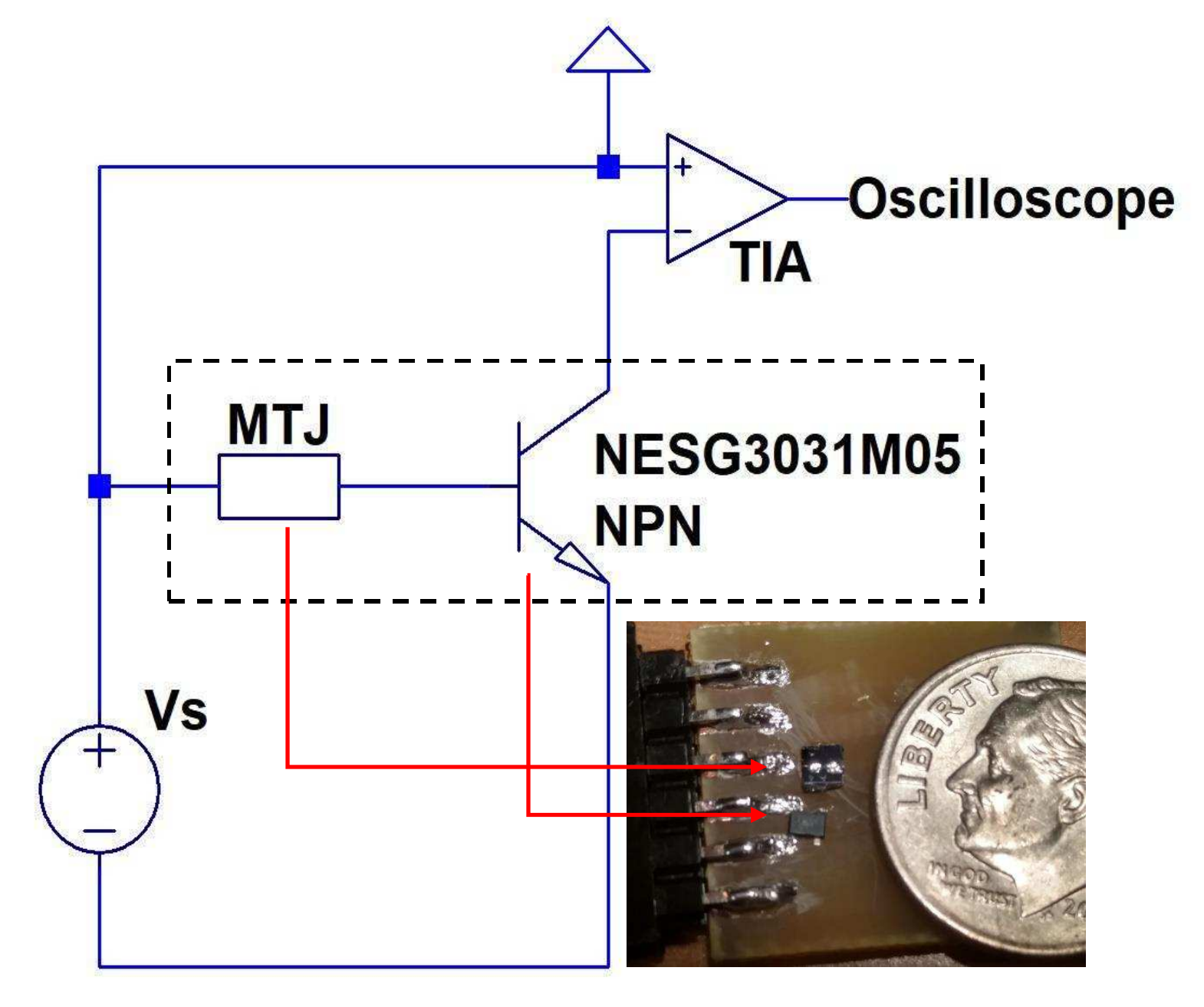}
\caption{\label{circuitFig}Circuit used for measurement. Dashed box is located at 8K inside the cryostat. In the bottom right is a picture of the device and transistor mounted on the printed circuit board sample mount next to a dime for a size reference.}
\end{figure}

The MTJ is connected directly to the base of the SiGe HBT. As the magnetization of each thin film moves, the MTJ itself changes resistance. This change in resistance is monitored at the base of the transistor and leads to a change in collector current (see Fig.~\ref{circuitFig} for details). The resulting collector current is sent to an external, room temperature transimpedance amplifier (TIA). The output of the TIA is recorded by a 1~GHz oscilloscope. The results of the experiment show an increase in bandwidth, increase in SNR, and in some situations, even an effective gain of the TMR signal. The limitation is found to be in the room temperature TIA, instead of the actual MTJ-transistor system.

The article will first describe the MTJ fabrication process. After that, the screening process of the MTJs and SiGe HBTs will be discussed. Then, the experimental procedure for the compound MTJ-SiGe HBT system will be presented as well as the resulting measurement improvements. To finish, the magnetodynamcis that are revealed after implementing the changes in the measurement system will be reported.

\section{\label{fab}MTJ Fabrication}

\begin{figure}
\setlength{\abovecaptionskip}{5pt}
\setlength{\belowcaptionskip}{-10pt}
\includegraphics[scale=.46]{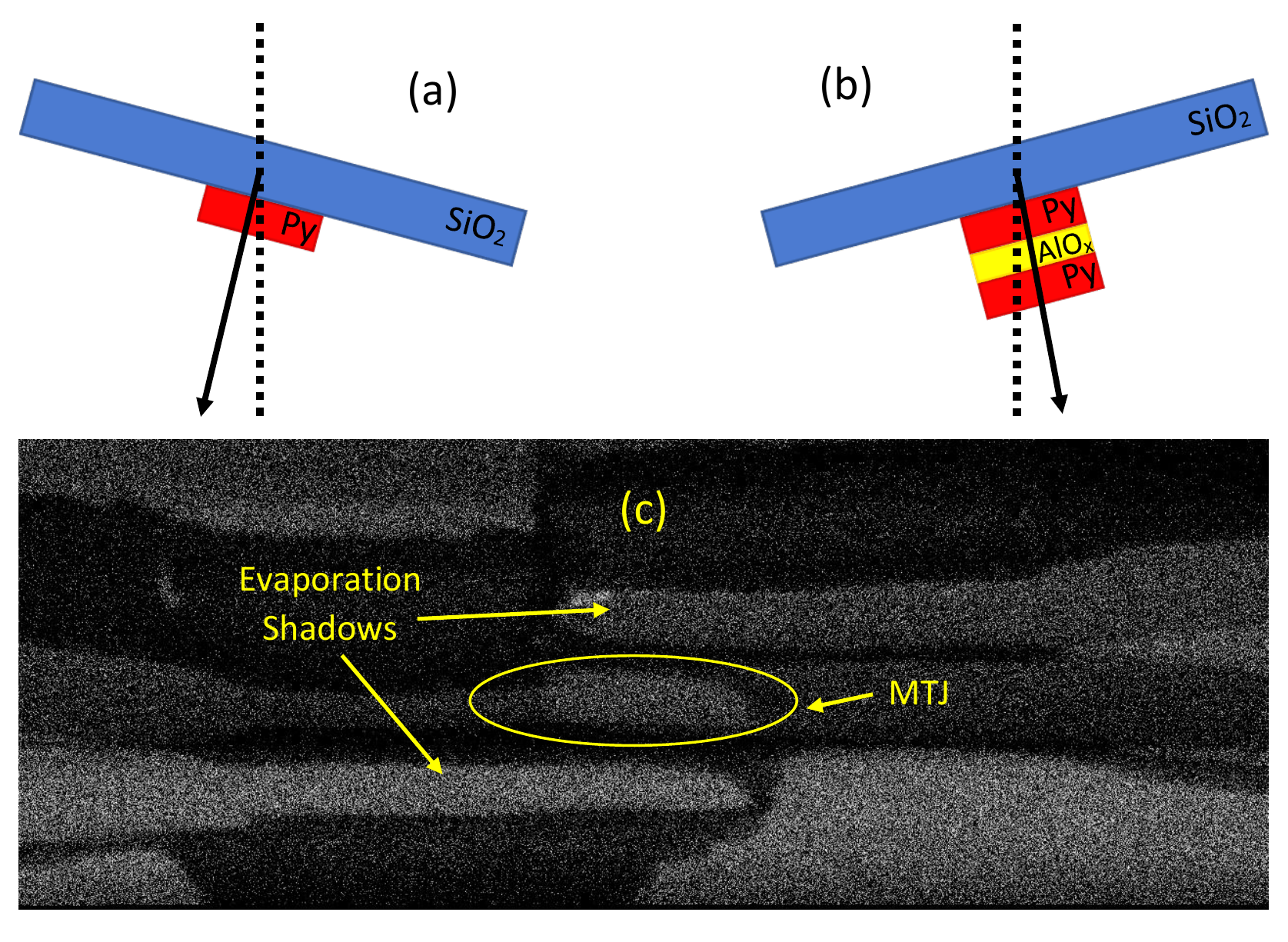}
\caption{\label{fabFig}(a) and (b) show the sample fabrication process (not to scale). The dotted line is the z-axis and extends down to the evaporation source. The PMMA bridge (not shown) and rotation of the sample platform allow for stacking of the materials to form the junction while simultaneously creating leads with connections to each permalloy layer. (c) is an SEM image of a junction made with chromium and gold with no oxide for imaging purposes only. The junction is in the center, and the two evaporation shadows are results of the process.}
\end{figure}

The fabrication of the MTJs begins with a 100 mm silicon wafer coated with a dry oxide. After the wafer is cleaned with acetone and isopropyl alcohol (IPA), it is then spin-coated with a layer of MMA followed by a layer of PMMA to create a bilayer electron beam resist. Pieces of the wafer are broken off for fabrication of arrays of twenty-four MTJs. Each separate chip is etched by electron beam lithography using a JEOL JSM-5910 Scanning Electron Microscope. Development is performed by vigorously stirring the sample in a 3:1 mixture of IPA and methyl isobutyl ketone (MIBK) for 60 seconds. Developing the bilayer electron beam resist forms a PMMA bridge over an empty cavity, on the silicon dioxide surface. This cavity allows for the stacking of materials by a process known as shadow evaporation deposition. The chip is then placed in a vacuum chamber and pumped down to 2.67E-5 Pa (2E-7 Torr) before deposition. Shadow evaporation deposition is performed using thermally heated materials and an Inficon Deposition Monitor to verify deposition rates with angstrom-level precision.

For shadow evaporation deposition, the stage is first tilted to form a 15$^{\circ}$ angle between the normal of the sample stage and the z-axis. The z-axis in this case extends from the evaporation source to the center of the sample stage as seen in Fig \ref{fabFig}. At this angle, 100 \AA{} of permalloy (Ni80/Fe20) is deposited. Next, the stage is tilted in the opposite direction making another 15$^{\circ}$ angle between the normal of the sample stage and the z-axis. The 35 \AA{} insulating barrier of alumina is deposited by evaporating aluminum at a rate of 2.5 \AA /s and releasing oxygen gas at the same time near the sample stage at a rate of 200 cm$^\text{3}$/min. This process forms alumina on the permalloy and silicon dioxide surfaces as the aluminum deposits. After the alumina, another 100 \AA{} of permalloy is deposited to finish the MTJ fabrication. 

Because the junction fabricated uses the same ferromagnetic material for each thin film, the width of each film was adjusted to ensure that the films in each junction have different switching fields. By having different switching fields, the magnetization of each film will flip at different fields giving the MTJs well-defined parallel and antiparallel states. Both layers have the same thickness as stated previously, but the bottom layer has a width of 300 nm and the top layer, 150 nm. The overlap of the two thin films forming the MTJ is 1 $\mu$m.
\section{\label{Screening}Device Screening and Experimental Setup}
The screening process for the MTJs as well as the SiGe HBTs is aimed at locating the devices with the optimal characteristics for the experiment. On the MTJ side, the goal is to find the junctions that have the largest absolute change in resistance during a magnetic field sweep at 8 K. For the transistors, the goal is to maximize the transconductance between the base and collector while staying within a stable range of operation. The screening procedures will be discussed next.

\begin{figure}
\setlength{\abovecaptionskip}{5pt}
\setlength{\belowcaptionskip}{-10pt}
\includegraphics[scale=.95]{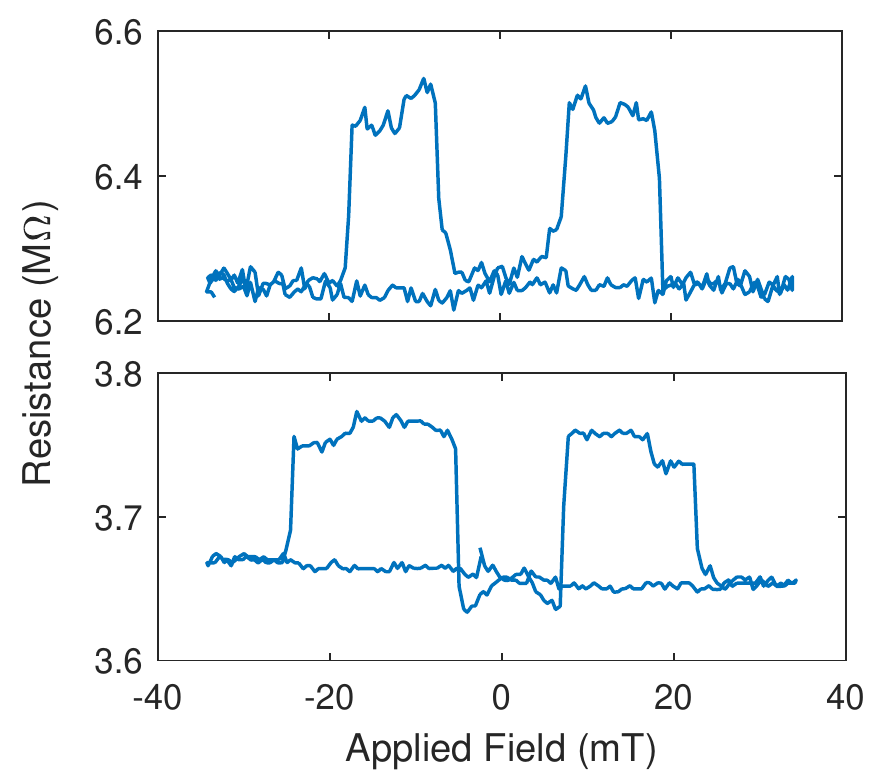}
\caption{\label{slowHystFig}Typical field sweeps of two different MTJs. The external field was applied as a triangle wave at a frequency of 0.1~Hz. The TIA rise time was set to 30~ms, and the data was acquired at a sample rate of 30~Hz. The output of the TIA is converted back into a resistance. The top (bottom) plot corresponds to what is later labeled as Sample 2 (1). }
\end{figure}

The MTJs are first screened at room temperature, by measuring the resistance of each junction. Samples with room temperature resistance down to about 10 k$\Omega$ were found to still produce TMR at 8 K; however, the absolute change in resistance was low. On the other hand, samples with room temperature resistance higher than 10 M$\Omega$ were also found to produce TMR; however, these samples had problems with excess noise. As a result, samples in the 1 M$\Omega$ range are chosen for further screening.

Once cooled to a base temperature of 8 K, magnetic field sweeps are performed to further screen samples for the largest absolute change in resistance. The external field is produced by a homemade superconducting solenoid mounted inside the cryostat. Using the geometry and winding density of the solenoid, the applied field per ampere of sourced current was calculated to be 27.7~mT/A. A Stanford Research Systems DS335 Function Generator produces a triangle wave that is fed into a Kepco BOP 20-20M power amplifier. The amplifier powers the solenoid to produce a magnetic field along the easy axes of the MTJs. The frequency of the triangle wave is set to 0.1~Hz, and the amplitude is adjusted so that the magnetic field amplitude is between 30~mT - 50~mT depending on the field needed to change the sample from the parallel to antiparallel then back to the parallel state. A constant voltage is supplied to the MTJs, and the current through the junction is monitored with an Ithaco Model 1211 TIA. The TIA rise time is set to 30~ms to filter out high frequency noise. The voltage output of the TIA is recorded using a NI BNC-2120 Data Acquisition Board and LabView at a sample rate of 30~Hz to avoid oversampling. Typical results of a field sweep on a good sample with the voltage output converted back to a resistance can be seen in Fig \ref{slowHystFig}.

\begin{figure}
\setlength{\abovecaptionskip}{5pt}
\setlength{\belowcaptionskip}{-10pt}
\includegraphics[scale=.62]{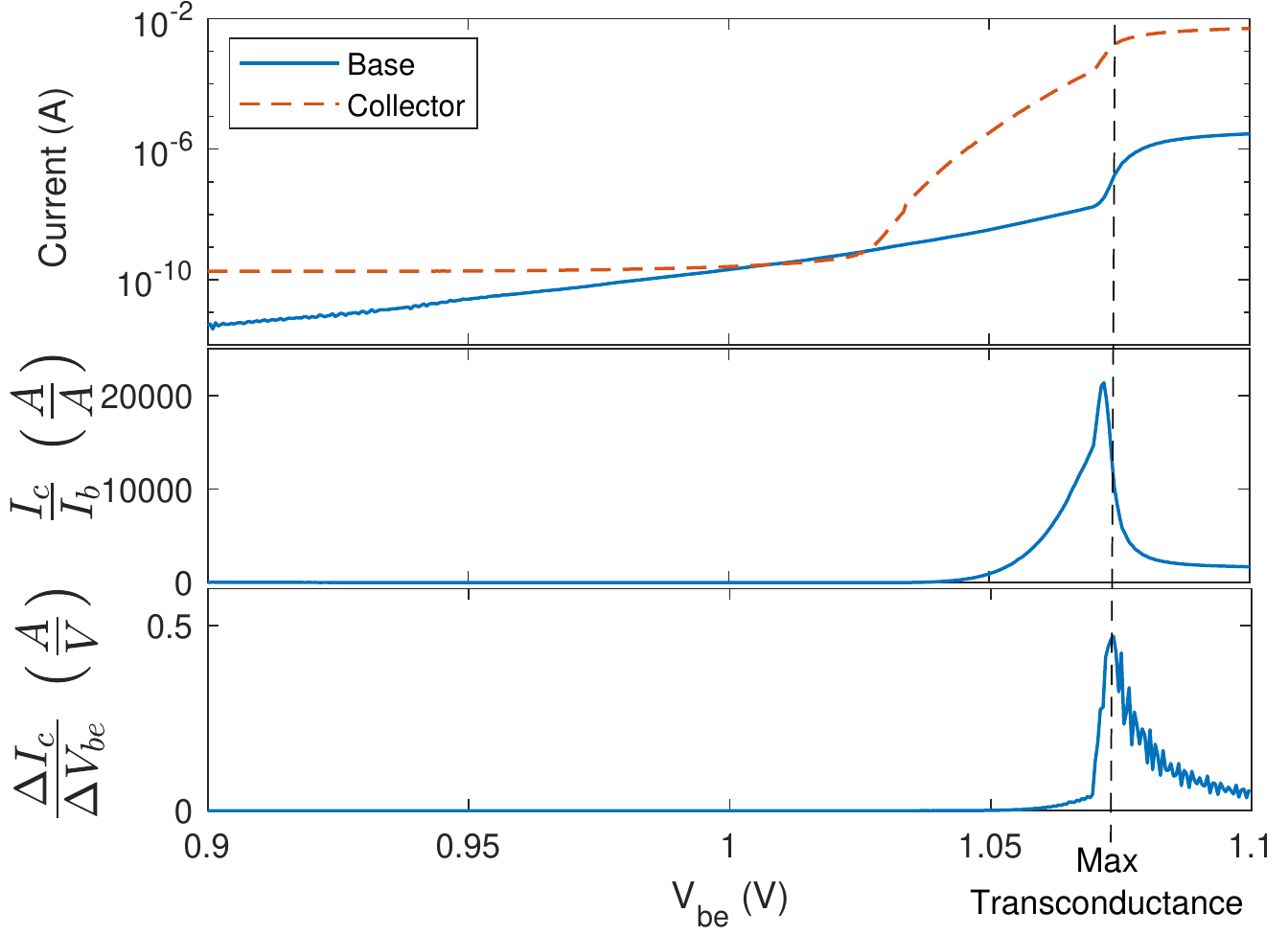}
\caption{\label{gummel}Top: Collector and base current versus the base-emitter voltage at a constant collector-base voltage(Gummel plot). Base 10 log scale is used for the y-axis. The collector and base are biased at the same voltage, and the current through each is measured. \newline Middle: DC Current gain of the output collector current and input base current versus the base-emitter voltage. \newline Bottom: Differential transconductance of the output collector current and the input base-emitter voltage vs the base-emitter voltage.  A vertical dashed line is drawn through the transconductance maximum to show that it drops off after the jump in the base current.}
\end{figure}

The SiGe HBT used for this experiment is an ``off-the-shelf'' CEL NESG3031M05 (NESG). At room temperature, these transistors are stable and pretty consistent from one device to another. Once cooled to 8 K, their operation begins to vary from sample to sample. To screen the transistors, the base and collector currents are measured with the TIA as the supply voltage is swept with a Stanford Research Systems DS345 Function Generator. The base and collector are both biased at the same voltage during the measurements allowing for the creation of a Gummel plot as seen in Fig \ref{gummel}.

At high bias ($>$1.06 V), a jump occurs in the transistor base current. This jump changes in height and position for each transistor. The jump is significant for two main reasons: First, due to the dimensions of the MTJs used, the samples can only withstand about a microamp of current before breakdown of the insulating barrier occurs resulting in permanent damage to the MTJ. If this jump raises the base current higher than this, the sample could blow, decreasing its resistance, and potentially damaging the transistor as well. Second, the gain of the transistor (both the DC current gain and the differential transconductance gain) continues to increase as the bias increases; however, after the jump occurs, the gain drops off (see Fig \ref{gummel}). This leads to the qualities of a ``good'' transistor as one that has the base current jump at a high bias, but still has the top of the jump in the microamp range. Throughout the screening process, approximately one out of every six transistors had optimal characteristics for the experiment.

Two other models of ``off-the-shelf'' SiGe HBT devices were also screened for possible use. These included the Infineon BFP720  and the NXP BFU710F. Both of these transistors operated at 8 K; however, both also had the same issue with the jump in the base current. In the end, the NESG was chosen due to the DC current gain right before the base current jump. As seen in Fig \ref{gummel}, the DC current gain of the NESG increases up to around 20,000. At the same point in the other transistor gummels, the gain only reached around a couple hundred. Considering the wide availability of SiGe HBTs today, future experiments will involve testing of additional device models as well. 

Once a ``good'' sample and ``good'' transistor are found, both devices are mounted onto a PCB with the base of the transistor as physically close to the MTJ as possible to reduce the input parasitic capacitance at the transistor base (see Fig. \ref{circuitFig}). Indium is used to bond small copper wires to connect the sample, transistor, and the printed circuit board (PCB) sample mount. The PCB is mounted inside the crysotat and connected to phosphor-bronze wires that exit the cryostat through a Fischer Connector. The wires are about 1~m long, and twisted together to reduce the effect of externally induced electrical noise. The entire system is cooled to 8K before performing further measurements. 

Initially, the current of the base and collector are each measured using a constant voltage bias across the collector and the sample-base connection while running through the same field sweep described previously. If an appreciable signal is seen, the TIA is connected to the collector, and the rise time is set to the minimum setting. The TIA output is sent directly to a Lecroy 9370 1GHz Oscilloscope to record the dynamics of the system as shown in Fig. \ref{circuitFig}. The frequency of the field sweep is also increased to allow for better time resolution of the measurement. 
\section{\label{results}Results and Discussion}
The results to be discussed can be categorized into two major sections: First, those that pertain improvements made to the measurement system will be discussed. These improvements include increased bandwidth, increased SNR, and an effective TMR signal gain. After, as a result of the measurement improvements, observed magnetodynamics of the system will be presented and discussed. The two MTJs used for the following analyses will be denoted as ``Sample 1'' and ``Sample 2'' and have room temperature resistances of 403 k$\Omega$ and 1.15 M$\Omega$, respectively.

\begin{figure}
\setlength{\abovecaptionskip}{5pt}
\setlength{\belowcaptionskip}{-10pt}
\includegraphics[scale=.62]{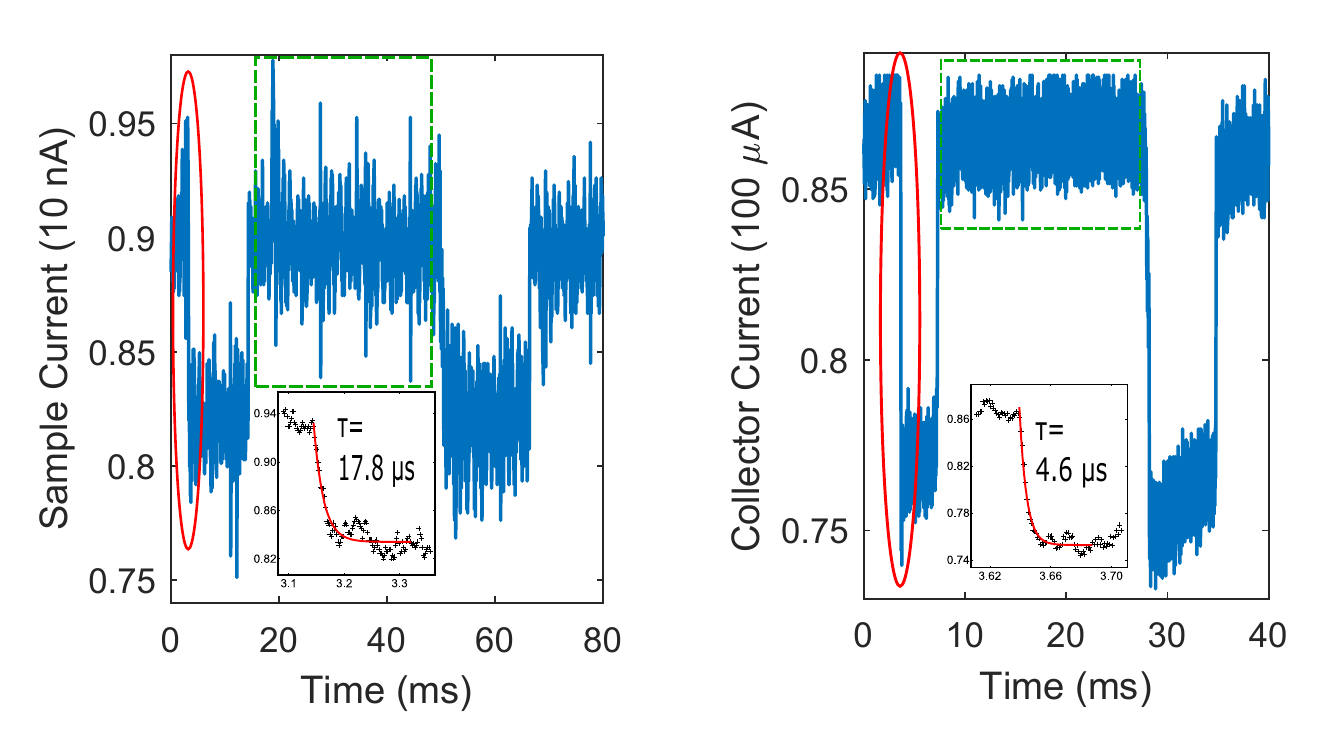}
\caption{\label{resultsE1}A current vs time plot of a broadband, single-shot field sweep of Sample 1 without(with) the transistor is on the left(right). The left(right) plot was taken with a field sweep rate of 10 Hz(20 Hz). The insets show data point markers and exponential fits of the first transition of each plot (in solid red ellipses) as well as the extracted time constant for each. The inclusion of the SiGe HBT decreases the time constant of the measurement by 74.3\%, from 17.8~$\mu$s to 4.57~$\mu$s, and increases the SNR by a factor of 6.62, from 6.48 to 42.9.}
\end{figure}

To get a good comparison for the bandwidth and SNR, a single-shot, broadband measurement of the MTJ's magnetic field response needs to be compared in the two configurations, with and without the SiGe HBT. The field response of interest is the change in resistance of the MTJ which is directly related to the configuration of the magnetizations of the permalloy thin films. Since resistance is a passive parameter, it makes more sense to compare the current through the MTJ at a constant applied voltage to the collector current of the SiGe-MTJ system. The initial broadband measurement of the MTJ by itself (without the SiGe HBT) proved difficult due to the small currents involved being plagued with too much noise to observe any field dependent dynamics throughout the field sweep. Of the two samples being discussed, Sample 1 had a large enough TMR to combat the external noise (see Fig.~\ref{resultsE1}). An exponential fit was applied to the first transition of each field sweep(inset of Fig.~\ref{resultsE1}) to find the time constant which can be directly related to the bandwidth of the measurement by Eq.~\ref{bandwidth}.
\begin{equation}
f_{bw}=\frac{1}{2\pi{}\tau{}}
\label{bandwidth}
\end{equation}
Where $\tau$ is the measured time constant, and $f_{bw}$ is the bandwidth. 

The measured time constant without the transistor came out to be 17.8 $\mu$s which corresponds to a bandwidth of 8.94 kHz. After Sample 1 was connected to the transistor and cooled, a single shot, broadband measurement of the collector current was recorded. The time constant was found to be 4.57 $\mu$s which corresponds to a bandwidth of 34.8 kHz. This 74.3\% decrease in time constant (or an increase in bandwidth by a factor of 3.89) gives preliminary evidence that including the transistor speeds up the measurement. Even better, the average resistance of Sample 1 after cooling was 3.67 M$\Omega$ which leads to an effective parasitic capacitance of only 1.25pF. 

The second measurement (right side of Fig. \ref{resultsE1}) was taken with an increased field sweep frequency (20 Hz instead of 10 Hz). The frequency was increased to allow for better time resolution in the measurement, but one could argue that this change in experimental conditions voids the conclusion. If we instead look at the specifications given for the TIA, we find that different gain settings have different minimum 10\% to 90\% rise times. Assuming a simple RC circuit, the time constant can be related to the 10\% to 90\% rise time by Eq.~\ref{risetime}.
\begin{equation}
\tau{}_{RC}=\frac{\tau{}_{rt}}{ln(9)}
\label{risetime}
\end{equation}
Here $\tau{}_{RC}$ is the RC time constant of the filter, and $\tau{}_{rt}$ is the 10\% to 90\% rise time. For the first and second measurement, the minimum rise times for the gain settings are listed as 40 $\mu$s and 10 $\mu$s, respectively. These rise times correspond to time constants of 18.2 $\mu$s and  4.55 $\mu$s. The closeness of these numbers to our measured time constants suggests that the TIA is limiting the speed of the measurement. As a result, incorporating the transistor leads to a larger current being sent to the TIA. The larger current allows for a smaller gain to be used, and the smaller gain setting allows for a faster measurement. From this analysis, including the transistor does, in fact, increase the overall bandwidth of the measurement system.

The SNR is the ratio of the signal power to the noise power. Because the measurements both have DC offsets, the applicable signal and noise currents can be expressed by the standard deviation of each. The signal, as stated previously, is a current measurement either through the MTJ (left of Fig.~\ref{resultsE1}) or through the collector of the SiGe HBT (right of Fig.~\ref{resultsE1}), and includes the evolution of the MTJ from the parallel to the anti-parallel back to the parallel state. Higher currents occur in the parallel state and lower currents in the anti-parallel state. The noise can be isolated in a stationary part of the measurement (green boxes in Fig.~\ref{resultsE1}). Because these values are currents they need to be converted into powers. Since the signal and noise current go through the same impedance, the ratio of the two standard deviations squared will be the same as the ratio of the two powers (the impedance will cancel out). The reduced equation for the SNR comes out to be:
\begin{equation}
SNR=(\frac{\sigma{}_{signal}}{\sigma{}_{noise}})^{2}
\label{SNR}.
\end{equation}
Where $\sigma$ is used as the standard deviation for each set of values.

The resulting noise standard deviation, signal standard deviation, and SNR of the configuration without the transistor (left side of Fig.~\ref{resultsE1}) are 0.144~nA, 0.366~nA, and 6.48, respectively. Following the same process for the configuration with the transistor (right side of Fig.~\ref{resultsE1}) gives values of 626~nA for the noise, 4.10~$\mu$A for the signal, and 42.9 for the SNR. By incorporating the transistor in the measurement, the SNR was increased by by a factor of 6.62.

\begin{figure}
\setlength{\abovecaptionskip}{5pt}
\setlength{\belowcaptionskip}{-10pt}
\includegraphics[scale=.62]{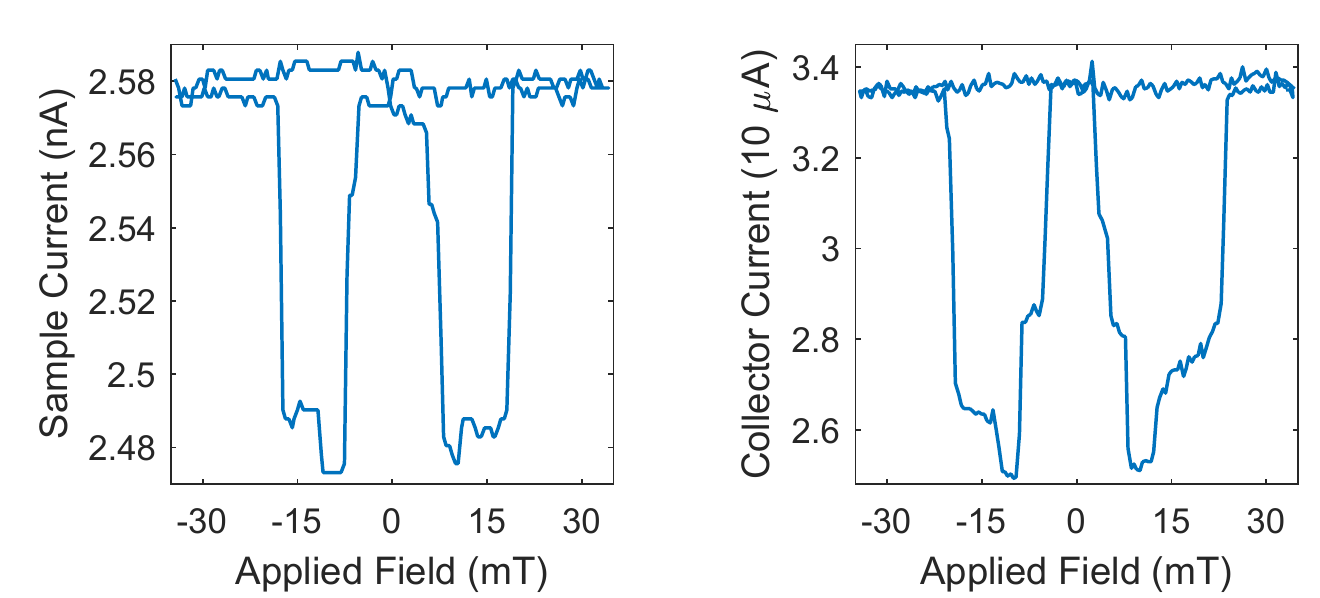}
\caption{\label{resultsC4}A current vs field plot of a field sweep of Sample 2 without(with) the transistor is on the left(right). The external field was applied as a triangle wave at a frequency of 0.1 Hz. The TIA rise time was set to 30ms, and the data was acquired at a sample rate of 30 Hz. The plots show the current through the MTJ and the collector of the SiGe HBT for easier comparison of the two configurations. The resulting TMR using Eq.~\ref{TMR} shows an increase by a factor of 7.75, from 4.45\% to 34.5\%, by adding the SiGe HBT to the measurement circuit.}
\end{figure}

\begin{figure*}
\setlength{\abovecaptionskip}{5pt}
\setlength{\belowcaptionskip}{-10pt}
\includegraphics[scale=.6]{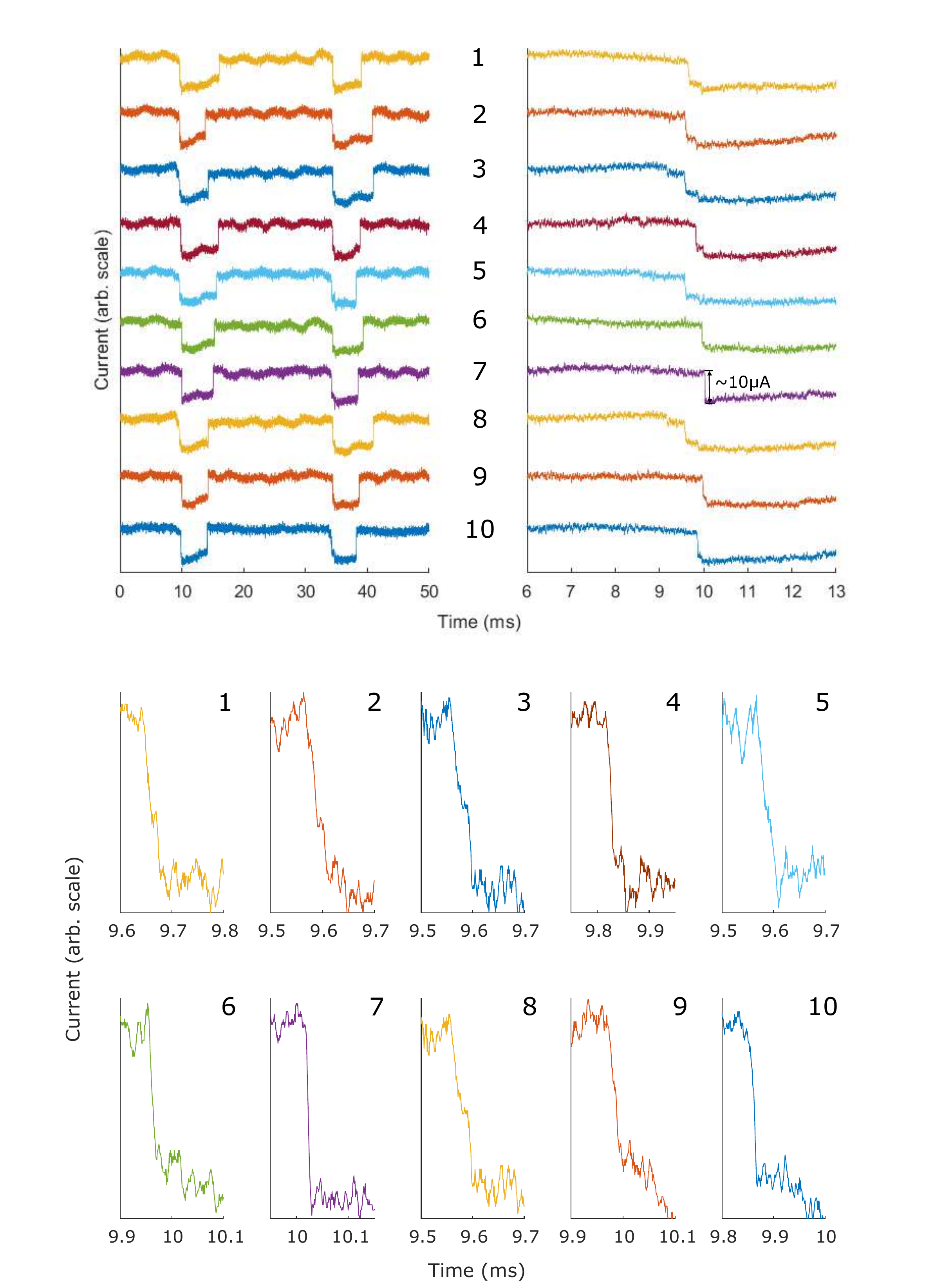}
\caption{\label{E1Dynamics}Plots of the SiGe HBT collector current vs. time while connected to Sample 1 during a field sweep. The applied field is at -19.3~mT at t=0~ms increasing at a rate of 3.51~T/s. The top left shows 10 consectutive traces of the 70 total traces throughout one period of the 20~Hz triangle wave of the applied field. The top right plot is of the same traces enlarged around the first transition (increasing field from the parallel to anti-parallel state). The bottom plots are greater enlargements of each transition corresponding to its given number. Although the transitions initially look discrete, with better time resolution, intermittent states on the order of tens of microseconds can be seen during the transition.}
\end{figure*}

For the effective signal gain, Sample 2 was chosen for analysis because it has the largest absolute change in resistance. The larger the change in resistance, the larger the change in the base-emitter voltage. The larger the change in base-emitter voltage, the larger the change in the collector current (the measured signal). Because the signal of interest is actually the TMR, to calculate the effective signal gain, the TMR of each configuration must be calculated. After the transistor is added, it is more applicable to discuss TMR in terms of the changes in current which gives rise to Eq.~\ref{TMR}
\begin{equation}
TMR=\frac{I_{P}}{I_{AP}}-1
\label{TMR}.
\end{equation}
Where $I_{P}$ and $I_{AP}$ refer to the current flow through the MTJ and collector of the transistor when the MTJ is in the parallel and anti-parallel state, respectively. As stated previously, the higher currents relate to the parallel state and the lower currents to the anti-parallel state.

For the configuration without the transistor (left side of Fig.~\ref{resultsC4}), the anti-parallel orientation current minimum is 2.47~nA, and the parallel current is about 2.58~nA. The resulting TMR is 4.45\%. For the configuration with the transistor, the amplified anti-parallel current minimum is 24.9~$\mu$A, and the amplified parallel current is about 33.5~$\mu$A resulting in an effective TMR of 34.5\%. By incorporating the transistor into the measurement setup, an effective TMR signal gain of 7.75 was achieved.

Although more improvements can be made to the system, adding a SiGe HBT to the measurement circuit inside the cryostat has improved the overall measurement process. The measurement bandwidth was increased, the SNR was increased, and, depending on the sample, the TMR signal can be increased as well.

To study the magnetodynamcis of the MTJs, multiple traces were acquired of each sample interfaced with the SiGe HBT and analyzed. For measurements, the field sweep rate was set to 20 Hz as stated previously. The TIA was connected to the collector of the SiGe HBT and its output went to the oscilloscope (see Fig. \ref{circuitFig}. The oscilloscope was set to trigger during the increasing portion of the magnetic field. The resulting traces on the oscilloscope show the collector current in respect to time and directly relate the time-resolved magnetization of the MTJ to specific applied fields. These traces show variations in the switching field, which has been studied previously\cite{Wernsdorfer1997}, and, more importantly, features that were unresolved before adding the SiGe HBT. There were 70 single-shot traces captured of Sample 1 and 30 single-shot traces of Sample 2.

For Sample 1, the applied field at t=0~ms was -19.3~mT and increasing at a rate of 3.51~T/s. Sample 1 appears to have a discrete transition between the parallel and anti-parallel configurations according to Figs.~\ref{slowHystFig}~and~\ref{resultsE1}. A closer look at the first transition between the parallel and anti-parallel states reveals that 64 out of the total 70 traces have steps during this transition. Fig.~\ref{E1Dynamics} shows ten of the total traces with two separate magnifications of the first transition. The traces are plotted with an offset in the y-axis for visibility purposes; however, a current-scale is added to one of the discrete transitions as a reference. Once the time range was magnified to 200~$\mu$s, a small step could be seen during the transition. The average lifetime of this step was only 23.4~$\mu$s; however, the traces that do not appear to have any steps in the transition could have steps that are still unresolved. 

\begin{figure}
\setlength{\abovecaptionskip}{5pt}
\setlength{\belowcaptionskip}{-10pt}
\includegraphics[scale=.8]{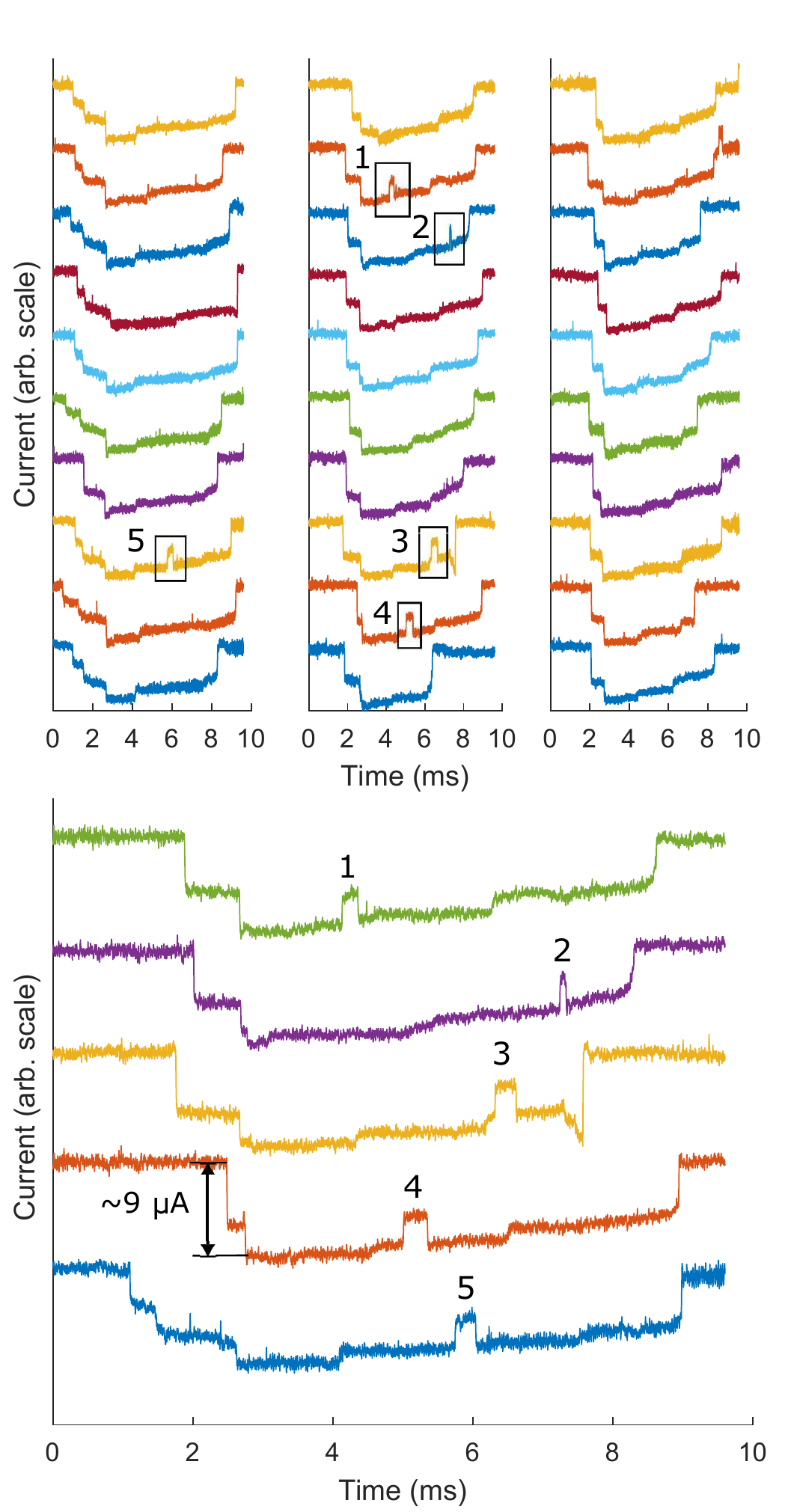}
\caption{\label{C4Dynamics}Plots of the SiGe HBT collector current vs. time while connected to Sample 2 during a field sweep. The applied field is at 7.45~mT at t=0~ms increasing at a rate of 2.72~T/s. The top plots are of all 30 collected traces, and the bottom plots are an enlarged view of the five traces that had transitions during the anti-parallel state. These transitions appear to return to one of the previous intermittent resistance states.}
\end{figure}

Before connecting the SiGe HBT, the high bandwidth, single shot measurement of Sample 2 was too noisy to even see any TMR. After adding the SiGe HBT, 30 field sweeps were recorded. The applied field at t=0~ms was 7.45~mT and increasing at a rate of 2.72~T/s. All 30 traces are plotted in Fig.~\ref{C4Dynamics} with an offset on the y-axis for each trace for visibility purposes. A current-scale was added to trace number 4 for reference. Looking at the traces, multiple steps during the transitions between the parallel and anti-parallel states of the MTJ can be seen, as well as a few transitions in the midst of the anti-parallel state. The average lifetime of these anti-parallel state transitions across the five traces in which they occur is 251~$\mu$s. In contrast to Sample 1, the steps seen in the transition from the parallel to anti-parallel states in this junction are seen in the slower field sweeps of Figs.~\ref{slowHystFig} and \ref{resultsC4}. Also, transitions during the anti-parallel state seen in this junction are not seen in Sample 1 at all. These two observations suggest slightly different mechanisms at play throughout the switching of these two junctions.

In both Sample 1 and Sample 2, the permalloy thin films that make up each MTJ most likely form multiple domains. This is due to both the size and shape of each thin film, a subject studied in previous literature\cite{Kirk1997}. This fact in addition to the stochastic nature of the effects seen in each sample suggest that the formation of the domains in each thin film during each field sweep plays a large role in the dynamics seen in each trace. 

In Sample 1, there are no transitions seen during the anti-parallel state reverting back to other previous resistance states which suggests that the transitions seen are ultimately irreversible. Irreversible sharp transitions during magnetic reversal have been seen before and are known as Barkhausen jumps\cite{Yang2005}. This would suggest that the magnetic reversal dynamics are dominated by domain wall motion in Sample 1. 

For Sample 2, the appearance of steps in the parallel to anti-parallel transition in the slow field sweep suggests that these resistance states are stable orientations of the domains of each thin film; however, the true stability of these states were not directly tested. In addition, the significant transitions seen during the anti-parallel state support this idea of stable domain orientations. Looking at the magnified traces at the bottom of Fig.~\ref{C4Dynamics}, four out of the five traces appear to revert back to a previous step in the first transition. Trace number 3, appears to transition to the center of a previous state; however, it is possible that due to the variation in switching fields, that two steps occurred very close to each other appearing as one larger step. These observations suggest that the magnetic reversal of Sample 2 is at least in part due to the rotation of domains within the permalloy thin films. This would allow domains to rotate between different stable orientations leading to reversible, intermittent resistance states.

\section{\label{sec:level1}Conclusion}

Interfacing SiGe HBTs with MTJs increases the bandwidth and SNR of the measurement. Depending on the sample, this measurement system can also create an effective TMR gain. The bandwidth limitation was shown to be from the external equipment which gives opportunity for further improvements to the system. 

At this point, the measurement improvements already allowed for electron transport measurements of MTJs to reveal time-resolved magnetodynamics of the nanomagnet thin films that form each MTJ. Different magnetic reversal mechanisms appeared in two similar samples both of which were dependent on how the domains formed in each thin film during an applied magnetic field sweep.

A similar approach for improving cryogenic measurements has been used in other low temperature systems, but has the added benefit of potentially increasing the TMR signal in a magnetic system. This same cryogenic preamplification could possibly be used for future cryogenic magnetic systems including the study of ferromagnetic nanoparticles and cryogenic memory systems.

\begin{acknowledgments}
This research was supported in part (J. Dark, G. Nunn, and D. Davidovi\'c) by DOE contract DE-FG02-06ER46281, and (H. Ying and J. D. Cressler) by the Laboratory Directed Research and Development (LDRD) Program at Sandia National Laboratories (operated by NTES of Sandia, a wholly owned subsidiary of Honeywell International Inc., for DOE’s NNSA under contract DE-NA-0003525). 
\end{acknowledgments}

\bibliographystyle{aipnum4-1}
\bibliography{ref4}

\end{document}